\documentclass[lettersize,journal]{IEEEtran}
\usepackage{mathrsfs}
\usepackage{url,times,amsmath,color,amssymb,graphicx,epsfig,psfig,cite,psfrag,subfigure,dsfont,booktabs}
\usepackage{algorithm}
\usepackage{algorithmic}
\usepackage{bm}
\usepackage{float}
\usepackage{microtype}
\newtheorem{lemma}{Lemma}

\makeatletter

\newcommand{\Rmnum}[1]{\expandafter\@slowromancap\romannumeral #1@}
\makeatother

\ifCLASSINFOpdf

\else
\fi
\usepackage{cite}

\begin{document}

\title{\huge Optimizing Rate-CRB Performance for Beyond Diagonal Reconfigurable Intelligent Surface Enabled ISAC}

\author{Xiaoqi Zhang,~\IEEEmembership{Member,~IEEE}, Liang Liu,~\IEEEmembership{Senior Member,~IEEE}, Shuowen Zhang,~\IEEEmembership{Senior Member,~IEEE}, \\Weifeng Zhu, Haijun Zhang,~\IEEEmembership{Fellow,~IEEE}

\thanks{

This work was supported in part by the National Key Research and Development Project of China under Grant 2022YFB2902800; in part by the National Natural Science Foundation of China under Grant 62471421. (\emph{Corresponding author: Shuowen Zhang.})

X. Zhang, L. Liu, S. Zhang and W. Zhu are with the Department of Electrical and Electronic Engineering, The Hong Kong Polytechnic University, Hong Kong, China (e-mail: zhangxiaoqi@xs.ustb.edu.cn; \{liang-eie.liu; shuowen.zhang; eee-wf.zhu\}@polyu.edu.hk).

H. Zhang is with the Beijing Engineering and Technology Research Center for Convergence Networks and Ubiquitous Services, University of Science and Technology Beijing, Beijing 100083, China (zhanghaijun@ustb.edu.cn).
%

}}
\maketitle

\begin{abstract}
This letter considers a beyond diagonal reconfigurable intelligent surface (BD-RIS) aided integrated sensing and communication (ISAC) system, where the BD-RIS can help a multi-antenna base station (BS) serve multiple user equipments (UEs) and localize a target simultaneously. We formulate an optimization problem that designs the BS beamforming matrix and the BD-RIS scattering matrix to maximize UEs' sum rate subject to a localization Cramér-Rao bound (CRB) constraint and an additional unitary matrix constraint for the scattering matrix. Because unitary matrices form a manifold, our problem belongs to constrained manifold optimization. This letter proposes a log-barrier based Riemannian steepest ascent method to solve this problem effectively. Numerical results verify the effectiveness of our algorithm and the performance gain of the BD-RIS aided ISAC systems over the conventional RIS aided ISAC systems.
\end{abstract}


\IEEEpeerreviewmaketitle

\section{Introduction}
Integrated sensing and communication (ISAC) has been recognized as a pivotal technology for 6G wireless networks \cite{mag1}. In the absence of line-of-sight (LoS) paths to certain region, reconfigurable intelligent surfaces (RIS) can improve sensing range via steering signals to blind spots.  For localization, \cite{localization1,localization3} showed that the RIS can act as a passive anchor such that targets can be localized based on their distance and angle-of-arrivals (AOAs) to the RIS. For RIS-aided ISAC, \cite{RISISAC1} proposed a joint optimization of RIS phase shifts, radar beamforming, and base station (BS) transmission to ensure communication and sensing performance while suppressing interference, thereby reducing the required receiver dynamic range. \cite{RISISAC2} introduced a dual-RIS-aided ISAC system using a penalty dual decomposition  algorithm to boost both radar and communication performance. \cite{RISISAC3} addressed communication maximization under sensing constraints via fractional programming, majorization minimization, and alternating direction method of multipliers. 

Recently, a novel beyond diagonal RIS (BD-RIS) structure has garnered attention for its enhanced flexibility in scattering matrix design \cite{mag2}.  This letter aims to study the role of BD-RIS in 6G ISAC.  Note that BD-RIS assisted ISAC has  been investigated in \cite{BDRISISAC1,BDRISISAC2,BDRISISAC3}. In \cite{BDRISISAC1} and \cite{BDRISISAC2}, the localization metric is signal-to-noise ratio (SNR), whose relation to localization accuracy is unknown. Although \cite{BDRISISAC3} adopts the Cramér-Rao bound (CRB) as the localization metric, it assumes that a sensor is at the BD-RIS to directly estimate the target AOA, and the BD-RIS scattering matrix does not affect the localization performance.  In this letter, our objective is to study how to exploit the BD-RIS to improve the sensing and communication performance simultaneously.

To achieve the above goal, this letter considers a BD-RIS assisted ISAC system that consists of a multi-antenna BS, multiple single-antenna user equipments (UEs), an active target, and a BD-RIS, as shown in Fig. 1. We aim to design the BS beamforming matrix and the BD-RIS scattering matrix jointly for maximizing the UEs' sum rate, subject to the target localization CRB constraint, the BS's transmit power constraint, and the unitary matrix constraint about the BD-RIS scattering matrix \cite{BDRIS1}. The main contributions are as below.

\begin{itemize}
\item
Firstly, this letter proposes to utilize the alternating optimization (AO) technique to separately design either the BS beamforming matrix or the BD-RIS scattering matrix given the other one. The main challenge is to optimize the BD-RIS scattering matrix, which should be a unitary matrix. In this letter, we propose to utilize the adaptive Riemannian steepest ascent method \cite{RSD} to deal with this issue. This method updates the BD-RIS scattering matrix along the Riemannian gradient by mapping the unitary constraint to the stiefel manifold. 
\item
Secondly, the adaptive Riemannian steepest ascent method is applicable for problems just with a unitary matrix constraint. However, under our problem, besides the unitary matrix constraint, there is the target localization CRB constraint. To fit our problem into the framework of adaptive Riemmanian steepest ascent method, we make use of the log-barrier method to embed the localization constraint into the objective function. The solution satisfies both the localization and unitary matrix constraints.
\end{itemize}

\section{System Model and Problem Formulation}
\begin{figure}[t]
        \centering
        \includegraphics*[width=48mm,height=30mm]{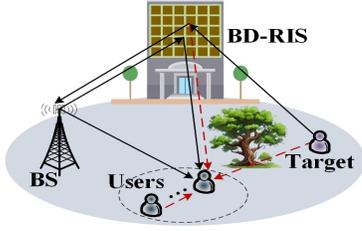}
        \caption{BD-RIS enabled ISAC system.}
        \label{fig:1}
\end{figure}
This letter considers a BD-RIS enabled ISAC system, which consists of a multi-antenna BS, $K$ single-antenna UEs, a single-antenna active target, and a BD-RIS, as shown in Fig. 1. Assume that the BS employs a full-duplex architecture with ${N_{{\mathop{\rm T}\nolimits} }}$ transmit antennas and ${N_{{\mathop{\rm R}\nolimits} }}$ receive antennas. It is assumed that the BS's transmit and receive antennas are sufficiently separated by 5-10 signal waveform to ensure that self-interference can be mitigated via radio frequency isolation and digital self-interference cancellation techniques \cite{self_interference}.

For the BD-RIS, we consider the group-connected structure \cite{BDRIS1}. Specifically, let ${\rm X}$ denote the number of groups, $M$ denote the total number of reflecting units, and ${M/{\rm{X}}}$ denote the group size. Then, the scattering matrix of the BD-RIS has the following block-diagonal structure \cite{BDRIS1}
 \begin{equation}
         {\rm{ }}{\bf{\Phi }} = {\rm{blkdiag}}( {{\bf{\Phi }}_1^{},...,{\bf{\Phi }}_\chi ^{},...,{\bf{\Phi }}_{\rm{X}}^{}} ),
\end{equation}
\begin{equation}
        {\bf{\Phi }}_\chi ^H{\bf{\Phi }}_\chi ^{} = {{\bf{I}}_{M/{\rm{X}}}},\forall \chi, 
 \end{equation}
where ${\bf{\Phi }}_\chi \in \mathbb{C}{^{{{M/{\rm{X}}}}\times {{M/{\rm{X}}}}}}$ denotes the scattering matrix of the $\chi$-th block of the BD-RIS. Note that each sub-block needs to be a unitary matrix due to the circuit requirement\footnote{BD-RIS is usually constrained by unitarity and symmetry, but the symmetry can be ignored with non-reciprocal circuits like circulators.}. The above model represents the conventional single connected mode when ${\rm X} = M$, and the fully connected mode when ${\rm X} = 1$.

In our considered system, the BS transmits downlink signals to the UE for communication purpose. Define ${{\bf{W}}_{{\rm{BS}}}} = [ {{{\bf{w}}_{{\rm{BS}},1}},...,{{\bf{w}}_{{\rm{BS}},K}}}]$, where ${{\bf{w}}_{{\rm{BS}},k}}$ denotes BS's transmit beamforming vector for UE $k$, and ${{\bf{s}}_{{\rm{BS}}}}[l] = {[ {{s_{{\rm{BS}},{\rm{1}}}}[l],...,{s_{{\rm{BS}},{{K}}}}[l]}]^T}$, where ${s_{{\rm{BS,}}k}}[l]$ denotes the communication symbol for UE $k$ at time slot $l$. At time slot $l$, the transmit signal is given as
\begin{equation}
        {\bf{x_{{\mathop{\rm BS}\nolimits}} }}[l] ={\bf W}{_{{\mathop{\rm BS}\nolimits} }}{{\bf s}_{{\mathop{\rm BS}\nolimits} }}[l].
\end{equation}
Moreover, the target transmits a signal to the BS for localization purpose. At time slot $l$, this signal is given as
\begin{equation}
        {x_{{\mathop{\rm Tar}\nolimits} }}[l] = \sqrt {{P_{{\mathop{\rm Tar}\nolimits} }}} {s_{{\mathop{\rm Tar}\nolimits} }}[l],
\end{equation}
where ${{P_{{\rm{Tar}}}}}$ is the transmit power of target, and ${s_{{\mathop{\rm Tar}\nolimits} }}[l]$ denotes the symbol of the target at time slot $l$.

\subsection{Sensing Model}
In this letter, we consider a challenging localization scenario, where the direct link between the target and the BS is blocked, as shown in Fig. 1. Then, based on the signals received by the BS over the target-RIS-BS link, we aim to estimate the AOA from the target to the BD-RIS, which is denoted as $\theta$. Let 
\begin{equation}
       {{\bf{r}}_{{\rm{Tar}}}}( {{\bf{\beta }},{\bf{\theta }}}) = \beta {\bf{a}}( {\bf{\theta }}),
\end{equation}
denote the channel between the BD-RIS and the target, where $\beta $ is the path loss, and ${\bf{a}}\left( {\bf{\theta }} \right)$ is the steering vector of the BD-RIS towards direction $\theta$. Moreover, let $L$ denote the number of time slots used by the BS for AOA estimation. At the $l$-th time slot, the signal received by the BS is
\begin{equation}
        {\bf{y}_{{\rm{TB}}}}[l] =  {{\bf{h}}_{{\rm{TB}}}}( {{\bf{\Phi }},{\bf{\beta }},{\bf{\theta }}}){x_{{\rm{Tar}}}}[l] + {\bf{n}_{{\mathop{\rm BS}\nolimits} }}[l],
\end{equation}
where 
\begin{equation}
        {{\bf{h}}_{{\rm{TB}}}}( {{\bf{\Phi }},{\bf{\beta }},{\bf{\theta }}}) = {\bf{G\Phi }}{{\bf{r}}_{{\rm{Tar}}}},
\end{equation}
denotes the cascaded channel of target-RIS-BS, ${\bf{G}} \in \mathbb{C} {^{{N_{{\mathop{\rm R}\nolimits} }} \times M}}$ denotes the channel of BS-RIS, and ${\bf{n}_{{\mathop{\rm BS}\nolimits} }}[l] \sim \mathcal{C} \mathcal{N} (0,\sigma _{{\mathop{\rm BS}\nolimits} }^2\bf{I})$ denotes the noise of the BS receive antennas at time slot $l$. 

Then the signals received over $L$ time slots is
\begin{equation}
        {{\bf{Y}}_{{\rm{TB}}}}\!\! =\! {[ {{\bf{y}_{{\rm{TB}}}}[1],...,{\bf{y}_{{\rm{TB}}}}[L]}]}\! = \!{{\bf{h}}_{{\rm{TB}}}}( {{\bf{\Phi }},{\bf{\beta }},{\bf{\theta }}} ){{\bf{x}}_{{\rm{Tar}}}^T} + {{\bf{N}}_{{\rm{BS}}}},
\end{equation}
where ${\bf{x}_{\rm Tar}} = [{{x}_{\rm Tar}}[1],...,{{x}_{\rm Tar}}[L]]^T$,  and ${\bf{N}_{\rm BS}} = [{\bf{n}_{\rm BS}}[1],...,{\bf{n}_{\rm BS}}[L]]$. 

Define ${\bm{\eta }} \buildrel \Delta \over = {[ {\theta ,{\bm{\beta }}^T} ]^T}$ as the collection of unknown variables, where ${\bm{\beta }} = {[ {\Re \left\{ \beta  \right\},\Im \left\{ \beta  \right\}} ]^T}$.
Based on \cite{FIM}, the Fisher information matrix (FIM) of ${\bm{\eta }}$  is
\begin{equation}
        {{\bf{\bar J}}_{\bm{\eta }}} = \left[ {\begin{array}{*{20}{c}}
        {{{{\rm{J}}}_{\theta \theta }}}&{{{\bf{J}}_{\theta \bm{\beta } }}}\\
        {{\bf{J}}_{\theta \bm{\beta } }^T}&{{{\bf{J}}_{\bm{\beta } \bm{\beta } }}}
        \end{array}} \right],
\end{equation}
where 
\begin{equation}
        {{\rm{J}}_{\theta \theta }} = \tfrac{{2{}L}}{{{\sigma _{{\rm{BS}}}^2}}}{\mathop{\rm tr}\nolimits} [ {{{{\bf{\dot h}}}_{{\mathop{\rm TB}\nolimits} }}( {\bf{\Phi }}){\bf{\dot h}}_{{\mathop{\rm TB}\nolimits} }^H( {\bf{\Phi }})} ],
\end{equation}
\begin{equation}
        {{\bf{J}}_{\theta \bm{\beta } }} = \tfrac{{2L}}{{{\sigma _{{\rm{BS}}}^2}}}\Re \{ {{\mathop{\rm tr}\nolimits}[ {{}{\bf{h}}_{{\mathop{\rm TB}\nolimits} }^{}( {\bf{\Phi }}){\bf{\dot h}}_{{\mathop{\rm TB}\nolimits} }^H( {\bf{\Phi }} )} ][1, \ {\rm{  }}j]} \},
\end{equation}
\begin{equation}
        {{\bf{J}}_{\bm{\beta } \bm{\beta } }} = \tfrac{{2L}}{{{\sigma _{{\rm{BS}}}^2}}}{\mathop{\rm tr}\nolimits}[ {{\bf{h}}_{{\mathop{\rm TB}\nolimits} }^{}( {\bf{\Phi }}){\bf{h}}_{{\mathop{\rm TB}\nolimits} }^H( {\bf{\Phi }} )} ]{{\bf{I}}_2},
\end{equation}
with ${\bf{\dot h}}_{{\mathop{\rm TB}\nolimits} }^{}( {\bf{\Phi }} )={{\partial ({\bf{h}}_{{\mathop{\rm TB}\nolimits} }^{}( {\bf{\Phi }}))} \mathord{\left/
 {\vphantom {{\partial ({\bf{H}}_{{\mathop{\rm Tar}\nolimits} }^{}\left( {\bf{\Phi }} \right))} {\partial \theta }}} \right.
 \kern-\nulldelimiterspace} {\partial \theta }}$.

The CRB matrix for estimating $\theta $ is 
\begin{equation}
        \begin{array}{l}
        {\rm{CR}}{{\rm{B}}_\theta }( {\bf{\Phi }}) = {[ {{\bf{\bar J}}_{\bm{\eta }}^{ - 1}} ]_{1,1}} = {{[ {{{\rm{J}}_{\theta \theta }} - {{\bf{J}}_{\theta {\bm{ \beta}} }}{{( {{{\bf{J}}_{{\bm{\beta}} {\bm{\beta}} }}})}^{ - 1}}{\bf{J}}_{\theta {\bm{ \beta}} }^T} ]}}^{ - 1}\\
        =\frac{{\sigma _{{\mathop{\rm BS}\nolimits} }^2}}{{2L{}\left( {{\mathop{\rm tr}\nolimits} \left( {{\bf{\dot h}}_{{\mathop{\rm TB}\nolimits} }^{}\left( {\bf{\Phi }} \right){\bf{\dot h}}_{{\mathop{\rm TB}\nolimits} }^H\left( {\bf{\Phi }} \right)} \right) - \frac{{{{\left| {{\mathop{\rm tr}\nolimits} \left( {{\bf{h}}_{{\mathop{\rm TB}\nolimits} }^{}\left( {\bf{\Phi }} \right){\bf{\dot h}}_{{\mathop{\rm TB}\nolimits} }^H\left( {\bf{\Phi }} \right)} \right)} \right|}^2}}}{{{\mathop{\rm tr}\nolimits} \left( {{\bf{h}}_{{\mathop{\rm TB}\nolimits} }^{}\left( {\bf{\Phi }} \right){\bf{h}}_{{\mathop{\rm TB}\nolimits} }^H\left( {\bf{\Phi }} \right)} \right)}}} \right)}}.
\end{array}
\end{equation}

\subsection{Communication Model}
Define the direct channel between the BS and the $k$-th UE as ${{\bf{d}}_{{\mathop{\rm BU}\nolimits,k} }}\!\! \in\! \mathbb{C}  {^{{N_{\rm T}}\times 1}}$,  and the channel between the BD-RIS and the $k$-th UE as ${{\bf{r}}_{{\mathop{\rm UE}\nolimits,k} }}\!\! \in \!\mathbb{C} {^{M \times 1}}$. Then, define the effective channel from the BS to the $k$-th UE that is contributed by both the direct link and the BD-RIS link as
\begin{equation}
{{\bf{h}}_{{\rm{BU}},k}}( {\bf{\Phi }} ) = {{\bf{d}}_{{\rm{BU}},k}} + {\bf{G\Phi }}{{\bf{r}}_{{\rm{UE}},k}}.
\end{equation}
The channels ${{\bf{d}}_{{\mathop{\rm BU}\nolimits,k} }}$, ${{\bf{r}}_{{\mathop{\rm UE}\nolimits,k} }}$, and ${\bf{G}}$ follow Rician fading with both LoS and non-line-of-sight (NLoS) components \cite{RISISAC3}, where the NLoS part follows a complex Gaussian distribution. This letter considers  a quasi-static block fading environment, where channels remain constant over each ISAC coherence interval.

Consider the interference from other UEs and the target, the downlink SINR of the $k$-th UE is expressed as
\begin{equation}
\begin{array}{l}
{\mathop{\rm \gamma }\nolimits} _{{\rm{UE}},k}^{}\!\!=\!\!
  {\textstyle{{{{\left| {{\bf{h}}_{{\rm{BU,}}k}^H\left( {\bf{\Phi }} \right){{\bf{w}}_{{\rm{BS,}}k}}} \right|}^2}} \over {\sum\limits_{i = 1,i \ne k}^K {{{\left| {{\bf{h}}_{{\rm{BU}},k}^H\left( {\bf{\Phi }} \right){{\bf{w}}_{{\rm{BS,}}i}}} \right|}^2}} {\rm{ + }}{{\left| {{{\rm{d}}_{{\rm{TU}},k}} + {\bf{r}}_{{\rm{UE}},k}^H{\bf{\Phi }}{{\bf{r}}_{{\rm{Tar}}}}} \right|}^2}{P_{{\rm{Tar}}}} + \sigma _{{\rm{UE}},k}^2}}},
\end{array}
\end{equation}
where ${{{\rm d}_{{\rm TU},k}}}$ denotes the direct channel from the target to the $k$-th UE, and $\sigma _{{\mathop{\rm UE}\nolimits,k} }^2$ is noise power at the $k$-th UE. Since the target transmits active signals to the BS for localization, if the condition $|{{\rm{d}}_{{\rm{TU}},k}} + {\bf{r}}_{{\rm{UE}},k}^H{\bf{\Phi }}{{\bf{r}}_{{\rm{Tar}}}}{|^2}{P_{{\rm{Tar}}}} > |{\bf{h}}_{{\rm{BU}},k}^H{{\bf{w}}_{{\rm{BS}},k}}{|^2} + \sum\nolimits_{i \ne k}^K {|{\bf{h}}_{{\rm{BU}},k}^H{{\bf{w}}_{{\rm{BS}},i}}{|^2}}  + \sigma _{{\rm{UE}},k}^2$ is satisfied, UE $k$ can subtract these known signals to enhance SINR. This depends on channel conditions, BD-RIS configuration, and BS beamforming.

Therefore, the downlink sum rate of the UEs is
\begin{equation}
R({{\bf{W}}_{{\rm{BS}}}},{\bf{\Phi }}) = \sum\nolimits_{k = 1}^K {{{\log }_2}(1 + \gamma _{{\rm{UE}},k}^{})}.
\end{equation}

\subsection{Problem Formulation}
In this letter, we aim to design the BS  beamforming matrix $\bf{W}_{\rm BS}$ and the BD-RIS scattering matrix ${\bf{\Phi }}$ to optimize the trade-off between the communication and the sensing performance. The problem is formulated as
\begin{subequations} 
        \begin{align}
        {\rm{}} (\rm{P}) \quad & \!\!\mathop {\max }\limits_{{{{\bf{W}}_{\rm{BS}}}},{\bf{\Phi }}} \ R({{{\bf{W}}_{\rm{BS}}}},{\bf{\Phi }}) \\
        &{\rm{s.t.}}  \qquad \!\!\! {\sum\nolimits_{k = 1}^K {{{\| {{{\bf{w}}_{{\rm{BS,}}k}}}\|}^2}}  \le {P_{\max }}},\\
        & \qquad \quad {{\rm{CRB}}_\theta }( {\bf{\Phi }} ) \le {\Delta _{\max }},\\
        & \qquad \quad (1), (2),
        \end{align}
\end{subequations}
where ${P_{\max }}$ and ${\Delta _{\max }}$ are the transmit power constraint and CRB  parameter estimation requirement, respectively.
\section{Proposed Solution}
Problem (P) is highly non-convex due to its complex objective function and CRB constraint, both of which are nonlinear, non-convex and difficult to handle. Moreover, the unitary constraint on the BD-RIS is non-convex, and the strong coupling between $\bf{\Phi}$ and $\bf{W}_{\rm BS}$ further adds to the challenge of joint optimization. To address this, we adopt an AO method. In the following, we show how to optimize $\bf{W}_{\rm BS}$ given ${\bf{\Phi }}$ and how to optimize ${\bf{\Phi }}$ given $\bf{W}_{\rm BS}$.
\subsection{BS Downlink Beamforming Design}
In this subsection, we show how to optimize $\bf{W}_{\rm BS}$ given ${\bf{\Phi }}$. Given any ${\bf{\Phi }}$,
the problem about ${{{\bf{W}}_{{\rm{BS}}}}} $ can be converted into 
\begin{equation} 
       {{\rm{}} (\rm{P1}) } \quad  \!\!\mathop {\max }\limits_{{{{\bf{W}}_{\rm{BS}}}}} \ R({{{\bf{W}}_{\rm{BS}}}}) \qquad
        {\rm{s.t.}}  \quad \!\!\! (17\rm b).
\end{equation}
According to the weighted minimum mean square error (WMMSE) framework in \cite{WMMSE}, Problem (P1) is equivalent to the following WMMSE minimization problem
\begin{equation} 
        {{\rm{}} (\rm{P1.1}) }  \!\!\mathop {\min }\limits_{\{ {u_k},{z_k},{\bf{w}}_{{\mathop{\rm BS}\nolimits} ,k}^{}\} } \sum\nolimits_{k = 1}^K {\left( {{z_k}{e_k} \!-\! \log {z_k}} \right)} \ \ 
        {\rm{s.t.}}  \quad \!\!\! (17\rm b),
\end{equation}
where ${z_k}$ denotes the weight of UE $k$. ${e_k}$ is the mean squared error of UE $k$, i.e., 
${e_k} \!\!=\!\! \mathbb{E} [ {{{| {{{\hat s}_k} - {s_k}} |}^2}} ]\!=\!\mathbb{E}[ {{{| {{u_k}{y_k} - {s_k}} |}^2}} ]$, and ${u_k}$ is the receiver coefficient for UE $k$.

With the other two parameters fixed, the optimal ${u_k}$ and ${z_k}$ are obtained by minimizing the objective of (P1.1), i.e.,
\begin{equation}
{u_k}\! \!=\!\! \tfrac{{{\bf{h}}_{{\rm{BU}},k}^H{{\bf{w}}_{{\rm{BS}},k}}}}{{\sum\nolimits_{i = 1}^K {{{| {{\bf{h}}_{{\rm{BU}},k}^H{{\bf{w}}_{{\rm{BS}},i}}} |}^{\rm{2}}}{\rm{ + }}{{| {{{\rm{d}}_{{\rm{TU}},k}} + {\bf{r}}_{{\rm{UE}},k}^H{\bf{\Phi }}{{\bf{r}}_{{\rm{Tar}}}}} |}^2}{P_{{\rm{Tar}}}} + \sigma _{{\rm{UE}},k}^2} }},
\end{equation}
\begin{equation}
{z_k} = {( {1 - u_k^*{\bf{h}}_{{\rm{BU}},k}^H{{\bf{w}}_{{\rm{BS}},k}}})^{ - 1}}.
\end{equation}
To satisfy constraint (17b), the Lagrangian dual method is used to derive the optimal solution of  ${\bf{w}}_{{\mathop{\rm BS}\nolimits} ,k}$, which is given by
\begin{equation}
{{\bf{w}}_{{\rm{BS}},k}}\!\! =\!\! {u_k}{z_k}{( {\omega {{\bf{I}}_{{N_{\rm T}}}} \!\!+\!\! \sum\nolimits_{i = 1}^K {{{| {{u_i}} |}^{\rm{2}}}{z_i}{\bf{h}}_{{\rm{BU}},i}^{}{\bf{h}}_{{\rm{BU}},i}^H} } )^{ - 1}}{\bf{h}}_{{\rm{BU}},k}^{},
\end{equation}
where $\omega >0$ denotes the optimal dual variable of the transmit power constraint. Then, ${\bf{w}}_{{\rm{BS}},k}$ is iteratively optimized according to the update rules of (20)-(22).

\subsection{BD-RIS Design}
Next, we show how to optimize ${\bf{\Phi }}$ given $\bf{W}_{\rm BS}$. Since (17c) is non-convex and difficult to handle, we first apply the Logarithmic barrier method \cite{covex} to transform Problem (P) with respect to ${\bf{\Phi }}$ into the following problem
\begin{subequations} 
        \begin{align}
        {\rm{}} (\rm{P2}) \quad & \!\!\mathop {\max }\limits_{\bf{\Phi }} \quad {\rm{ }}R ({\bf{\Phi }}) + {1 \mathord{/
 {\vphantom {1 \tau }}
 \kern-\nulldelimiterspace} \tau }\log ( {{\Delta _{\max }}  - {{\mathop{\rm CRB}\nolimits} _\theta }( {\bf{\Phi }} )} ) \\
        &{\rm{s.t.}}  \quad\quad \!\!\! (1), (2),
        \end{align}
\end{subequations}
where ${\tau >0}$ is a parameter that controls approximation accuracy. A larger value of $\tau$ weakens the CRB barrier’s influence, allowing the solution to move closer to the original problem. However, too large a $\tau$ can cause sharp gradients and instability. To manage this, we dynamically update $\tau$ using $\tau  = \nu\tau$, where $\nu$ is the decay factor. This balances stability, convergence speed, and solution quality. A smaller initial  $\tau$ ensures smooth optimization, while gradually increasing $\tau$ sharpens the barrier, guiding solutions toward feasibility and avoiding premature convergence.

However, directly solving Problem (P2) with conventional gradient-based methods is also challenging, as it’s hard to ensure each solution satisfies the unitary constraints. In this letter, we propose the adaptive Riemannian steepest ascent algorithm \cite{RSD} to desgin the BD-RIS scattering matrix with unitary constraints. Specifically, Problem (P2) is first mapped to a Riemannian manifold to remove constraints, then optimized along the corresponding Riemannian gradient.

First, for the $\chi$-th group of BD-RIS, the unitary constraint in Problem (P2) can be mapped onto a stiefel manifold, i.e.,
\begin{equation}
        {\mathcal{M}_\chi }= \{ {{\bf{\Phi }}_\chi \in \mathbb{C}{^{{{M/{\rm{X}}}}\times {{M/{\rm{X}}}}}}:{{\bf{\Phi }}_\chi^H}{\bf{\Phi }}_\chi = {\bf{I}}_{M/{\rm{X}}}} \}, \forall \chi.
\end{equation}

Next, we construct the Riemannian gradient of  $f({\bf{\Phi }})$ on manifold (24) and apply a geodesic-based search to avoid the extensive iterations of traditional methods.

Let 
$f({\bf{\Phi }}) \!=\! R({\bf{\Phi }}) \!+\! \frac{1}{\tau }\log ( {{\Delta _{\max }} - {\rm{CR}}{{\rm{B}}_\theta }( {\bf{\Phi }} )})$, adopting the geodesic strategy \cite{RSD}, the  Riemannian gradient is defined as 
\begin{equation}
{\bf{\Psi }}(f( {{{\bf{\Phi }}_\chi }})) = {\bf{\Sigma }}(f( {{{\bf{\Phi }}_\chi }} )){[ {{{\bf{\Phi }}_\chi }} ]^H} - {{\bf{\Phi }}_\chi }{[ {{\bf{\Sigma }}(f( {{{\bf{\Phi }}_\chi }} ))} ]^H},
\end{equation}
where ${\bf{\Sigma }}(f( {{{\bf{\Phi }}_\chi }}))$ denotes the the Euclidean gradient of $f({\bf{\Phi }}_\chi)$ with respect to ${\bf{\Phi }_\chi}$. According to the gradient solution rule for complex matrices \cite{qiutidu}, ${\bf{\Sigma }}(f( {{{\bf{\Phi }}_\chi }} ))$ is given by Lemma 1.
\begin{lemma} 
        Let $\rho ({\bf{\Phi }}_\chi) = {1 \mathord{\left/
        {\vphantom {1 \tau }} \right.
        \kern-\nulldelimiterspace} \tau }\log \left( {{\Delta _{\max }}  - {{\mathop{\rm CRB}\nolimits} _\theta }\left( {\bf{\Phi }}_\chi \right)} \right)$, ${\bf{\mathbf{U} }} = {\bf{h}}_{{\mathop{\rm TB}\nolimits} }^{}( {\bf{\Phi }}_\chi ){\bf{\dot h}}_{{\mathop{\rm TB}\nolimits} }^H( {\bf{\Phi }}_\chi )$, ${\bf{\mathbf{V} }} = {\bf{h}}_{{\mathop{\rm TB}\nolimits} }^{}( {\bf{\Phi }}_\chi ){\bf{h}}_{{\mathop{\rm TB}\nolimits} }^H( {\bf{\Phi }}_\chi )$, and $\xi  = {\rm{tr(}}{\bf{\dot h}}_{{\rm{TB}}}^{}( {\bf{\Phi }_\chi} ){\bf{\dot h}}_{{\rm{TB}}}^H( {\bf{\Phi }_\chi} ){\rm{)}} - \tfrac{{{{| {{\rm{tr}}( {{\bf{h}}_{{\rm{TB}}}^{}( {\bf{\Phi }_\chi} ){\bf{\dot h}}_{{\rm{TB}}}^H( {\bf{\Phi }_\chi} )} )} |}^2}}}{{{\rm{tr}}( {{\bf{h}}_{{\rm{TB}}}^{}( {\bf{\Phi }_\chi} ){\bf{h}}_{{\rm{TB}}}^H( {\bf{\Phi }_\chi} )} )}}$, then the Euclidean gradient of $f( {\bf{\Phi }}_\chi )$ is equivalent to
        \begin{equation}
            {\bf{\Sigma }}(f( {\bf{\Phi } }_\chi )) = {\bf{\Sigma }}(R( {\bf{\Phi }} _\chi)) + {\bf{\Sigma }}(\rho ( {\bf{\Phi }}_\chi )),
        \end{equation}     
where ${\bf{\Sigma }}(R( {{{\bf{\Phi }}_\chi }}))$ and ${\bf{\Sigma }}(\rho( {{{\bf{\Phi }}_\chi }}))$ denote the Euclidean gradient of $R( {{{\bf{\Phi }}_\chi }})$ and $\rho( {{{\bf{\Phi }}_\chi }})$ with respect to ${\bf{\Phi }_\chi}$, respectively, i.e.,
\begin{equation}
\begin{array}{l}
{\bf{\Sigma }(R({{\bf{\Phi }}_\chi }))} = \tfrac{1}{{\ln 2}} \times \\
(\sum\limits_{k = 1}^K {\tfrac{{{{( {{{\bf{M}}_1} + {{\bf{M}}_3}})}^T}}}{{\sum\limits_{i = 1}^K {{{\left| {{\bf{h}}_{{\rm{BU}},k}^H( {{{\bf{\Phi }}_\chi }} ){{\bf{w}}_{{\rm{BS,}}i}}} \right|}^2}} {\rm{ + }}{{| {{{\rm{d}}_{{\rm{TU}},k}} + {\bf{r}}_{{\rm{UE}},k}^H{{\bf{\Phi }}_\chi }{{\bf{r}}_{{\rm{Tar}}}}} |}^2}{P_{{\rm{Tar}}}} + \sigma _{{\rm{UE}},k}^2}}} -  \\
\sum\limits_{k = 1}^K {\tfrac{{{{( {{{\bf{M}}_1} + {{\bf{M}}_5}} )}^T}}}{{\sum\limits_{i = 1,i \ne k}^K {{{| {{\bf{h}}_{{\rm{BU}},k}^H( {{{\bf{\Phi }}_\chi }}){{\bf{w}}_{{\rm{BS,}}i}}}|}^2}} {\rm{ + }}{{| {{{\rm{d}}_{{\rm{TU}},k}} + {\bf{r}}_{{\rm{UE}},k}^H{{\bf{\Phi }}_\chi }{{\bf{r}}_{{\rm{Tar}}}}} |}^2}{P_{{\rm{Tar}}}} + \sigma _{{\rm{UE}},k}^2}}} ),
\end{array}
\end{equation}
        \begin{equation}
                        \begin{array}{l}
                       {\bf{\Sigma }}(\rho( {\bf{\Phi }}_\chi )) = \\
                        \frac{1}{{\tau ( {{{\mathop{\rm CRB}\nolimits} _\theta }( {\bf{\Phi }}_\chi )\! -\! {\Delta _{\max }}})}}( {\frac{{\sigma _{{\mathop{\rm BS}\nolimits} }^2}}{{2L{}{\xi ^2}}}}){[ {{{\bf{A}}_2} \!\!-\!\! \frac{{{{\bf{B}}_2}}}{{{\mathop{\rm tr}\nolimits} ( {\bf{V}} )}}\!\! +\!\!\frac{{{\mathop{\rm tr}\nolimits} ( {\bf{U}} ){{( {{\mathop{\rm tr}\nolimits} ( {\bf{U}} )})}^*}{{\bf{D}}_2}}}{{{{[ {{\mathop{\rm tr}\nolimits} ( {\bf{V}} )} ]}^2}}}} ]^T},
                        \end{array}
        \end{equation}
with
        \begin{equation}
                {{\bf{M}}_1}\!\! =\!\! {P_{{\rm{Tar}}}}( {{{\rm{d}}_{{\rm{TU}},k}}{\bf{r}}_{{\rm{UE}},k}^*{\bf{r}}_{{\rm{Tar}}}^T + {\bf{r}}_{{\rm{Tar}}}^*{\bf{r}}_{{\rm{Tar}}}^T{{\bf{\Phi }}_\chi ^T}{\bf{r}}_{{\rm{UE}},k}^*{\bf{r}}_{{\rm{UE}},k}^T} ),
                \end{equation}
                \begin{equation}
\begin{array}{l}
{{\bf{M}}_3} = {\bf{r}}_{{\rm{UE}},k}^*{\bf{d}}_{{\rm{BU}},k}^T\sum\nolimits_{i = 1}^K {{\bf{w}}_{{\rm{BS}},i}^*{\bf{w}}_{{\rm{BS}},i}^T} {{\bf{G}}^*}{\rm{ }} \\ \qquad  + {\bf{r}}_{{\rm{UE}},k}^*{\bf{r}}_{{\rm{UE}},k}^T{\bf{\Phi }}_\chi ^T{{\bf{G}}^T}\sum\nolimits_{i = 1}^K {{\bf{w}}_{{\rm{BS}},i}^*{\bf{w}}_{{\rm{BS}},i}^T} {{\bf{G}}^*},
\end{array}
                \end{equation}
                \begin{equation}
\begin{array}{l}
{{\bf{M}}_5} = {\bf{r}}_{{\rm{UE}},k}^*{\bf{d}}_{{\rm{BU}},k}^T\sum\nolimits_{i = 1,i \ne k}^K {{\bf{w}}_{{\rm{BS}},i}^*{\bf{w}}_{{\rm{BS}},i}^T} {{\bf{G}}^*} \\ \qquad  + {\bf{r}}_{{\rm{UE}},k}^*{\bf{r}}_{{\rm{UE}},k}^T{\bf{\Phi }}_\chi ^T{{\bf{G}}^T}\sum\nolimits_{i = 1,i \ne k}^K {{\bf{w}}_{{\rm{BS}},i}^*{\bf{w}}_{{\rm{BS}},i}^T} {{\bf{G}}^*},
\end{array}
                \end{equation}
                \begin{equation}
                        {{{\bf{A}}_2}} = {( {{\bf{G}}_{}^H{\bf{G}\bf{\Phi}_\chi\bf{\dot r}}_{{\mathop{\rm Tar}\nolimits} }^{}{\bf{\dot r}}_{{\mathop{\rm Tar}\nolimits} }^H} )^T},{{\bf{B}}_2} = {( {\bf{G}}_{{\mathop{\rm }\nolimits} }^H{{\bf{G}}{\bf{\Phi}_\chi}{\bf{\Omega }}} )^T},
                \end{equation}
                \begin{equation}
                        {{\bf{D}}_2} = {( {{\bf{G}}_{}^H{\bf{G}}{\bf{\Phi}_\chi}{\bf{ r}}_{{\mathop{\rm Tar}\nolimits} }^{}{\bf{r}}_{{\mathop{\rm Tar}\nolimits} }^H} )^T},
                \end{equation}
                \begin{equation}
                        {\bf{\Omega }} = {( {{\mathop{\rm tr}\nolimits} ( {\bf{U}} )} )^*}{{\bf{r}}_{{\mathop{\rm Tar}\nolimits} }}{\bf{\dot r}}_{{\mathop{\rm Tar}\nolimits} }^H + {\mathop{\rm tr}\nolimits} ( {\bf{U}}){{\bf{\dot r}}_{{\mathop{\rm Tar}\nolimits} }}{\bf{r}}_{{\mathop{\rm Tar}\nolimits} }^H.
                \end{equation}
\end{lemma} 

\newtheorem{proof}{Proof}
\begin{proof} 
       Please refer to Appendix A.
\end{proof} 

Like a straight line in Euclidean space, the geodesic in Riemannian space follows the shortest local path, defined by the exponential map, with the rotation matrix
\begin{equation}
{\bf{Q}_\chi} ={e^{  \mu {\bf{\Psi }}(f( {{{\bf{\Phi }}_\chi }} ))}}, 
\end{equation}
where $\mu \!\! > \!\!0$ is the step size. Since directly computing the matrix exponential is costly. To reduce complexity, efficient approximations like truncated Taylor series and diagonal Padé approximants \cite{RSD} are used.

 Using the rotation matrix, then ${\bf{\Phi }_\chi}$ is updated as 
\begin{equation}\label{18}
        \begin{array}{l}
{{{\bf{\Phi }}_\chi^{( {t + 1})}}} = {\bf{Q}_\chi} {{{\bf{\Phi }}_\chi^{( t )}}},
\end{array}
\end{equation}
where $t$ denotes the number of iterations.
\setcounter{algorithm}{0}
\renewcommand\baselinestretch{1}
\begin{algorithm}[t]\vspace{0pt}
\renewcommand{\thealgorithm}{1}
\caption{Joint Beamforming and BD-RIS Design for (P)}
\small
\begin{algorithmic}[1]
\STATE    Initialize  ${{\bf{\Phi }}^{\left( 0 \right)}}$, set  tolerance $\varepsilon, \epsilon$.
\REPEAT
\STATE    Update ${u_k}$, ${z_k}$, ${{\bf{w}}_{{\rm BS},k}^{(t)}}$ by (20), (21), (22).
\WHILE    {${{| {f( {{{\bf{\Phi }}^{({t_1})}}} ) - f( {{{\bf{\Phi }}^{({t_1} - 1)}}} )} |} \mathord{/
 {\vphantom {{| {f( {{{\bf{\Phi }}^{({t_1})}}} ) - f( {{{\bf{\Phi }}^{({t_1} - 1)}}} )} |} {f( {{{\bf{\Phi }}^{({t_1} - 1)}}} )}}} 
 \kern-\nulldelimiterspace} {f( {{{\bf{\Phi }}^{({t_1} - 1)}}} )}} > \varepsilon $}
\FOR      {$\chi=1 $ to $\rm X$}
\STATE    Update ${\bf{\Sigma }}(f( {\bf{\Phi } }_\chi ))$, ${\bf{\Psi }}(f( {{{\bf{\Phi }}_\chi }}))$, ${\bf{Q}}_{\chi}^{(t_1)}$ by (26), (25), (35).
\STATE    Calculate 
${\delta_\chi^{(t_1)}} =\frac{1}{2} \| {{\bf{\Psi }}(f( {{{\bf{\Phi }}_\chi^{(t_1)} }}))} \|_F^2$.
\WHILE   {$f ({\bf{Q}}_\chi^{(t_1)}{\bf{\Phi }}_\chi^{(t_1)}) - f  ({{\bf{\Phi }}_\chi^{(t_1)}}) < \frac{\mu }{2}{\delta_\chi^{(t_1)}}$}
\STATE    $\mu  = {\mu  \mathord{/
 {\vphantom {\mu  2}} 
 \kern-\nulldelimiterspace} 2}$. 
\ENDWHILE
\WHILE    {$f  ({\bf{Q}}_\chi^{(t_1)}{\bf{Q}}_\chi^{(t_1)}{\bf{\Phi }}_\chi^{(t_1)})\! - f  ({{\bf{\Phi }}_\chi^{(t_1)}}) \ge \mu {\delta_\chi^{(t_1)}} $}
\STATE    $\mu =2\mu $.
\ENDWHILE
\STATE    Update ${{\bf{\Phi }}_\chi^{(t_1+1)}}$ by (36).
\ENDFOR 
\IF       {$\tau  \le {1 \mathord{/
 {\vphantom {1 \varepsilon }} 
 \kern-\nulldelimiterspace} \varepsilon }$}
\STATE    Update $\tau  = \nu \tau$.
\ENDIF
\STATE    $t_1=t_1+1 $. 
\ENDWHILE
\STATE    $t=t+1 $. 
\UNTIL   $| {{R^{(t)}} - {R^{(t - 1)}}}|/{R^{(t - 1)}} \le \epsilon  $.
\end{algorithmic}
\end{algorithm}

The above process is illustrated in Algorithm 1. 
\subsection{Complexity Analysis}
For Problem (P1), the computational complexity is ${\cal O}\left( {{I_\omega }{I_{\bf w}}KN_T^3} \right)$, where $I_\omega $ and ${I_{\bf w}}$ are the number of iterations for optimizing $\omega$ and ${\bf w}$, respectively. For Problem (P2), the complexity of computing the Euclidean gradient and Riemannian gradient are ${\cal O}( {{K^2}N_T^{} \!+\! KM_\chi ^2N_T^{} \!+\! KN_T^2M_\chi ^{} \!+\! M_\chi ^3} )$ and ${\cal O}( {M_\chi ^3} )$, respectively. The complexity of computing the updated matrix and step size adaptation are ${\cal O}( {M_\chi ^3} )$ and ${\cal O}( {{I_\mu }M_\chi ^3} )$, respectively. Then the complexity of Problem (P2) is 
${\cal O}( {{I_{\bf{\Phi}}}{\rm X}( {{K^2}N_T^{} \!+\! KM_\chi ^2N_T^{} \!+\! KN_T^2M_\chi ^{} \!+\! {I_\mu }M_\chi ^3} )} )$, where  ${I_\mu }$ and  ${I_{\bf{\Phi}}}$ are the number of iterations for updating step size and optimizing ${\bf{\Phi}}$, respectively. Let ${I_{A}}$ denote the number of AO iterations, the overall complexity is  $
{\cal O}({I_{A}}({I_\omega }{I_{\bf{w}}}KN_T^3 \!\!+ \!\!{I_{\bf{\Phi }}}{\rm{X}}({K^2}N_T^{} \!\!+\!\! KM_\chi ^2{N_T} \!\!+\!\! KN_T^2M_\chi ^{}
 \!\!+ \!\!{I_\mu }M_\chi ^3))).
$


\section{Numerical Results}
\begin{figure}
	\centering
                \begin{minipage}[b]{0.49\linewidth}
                        \centering
                        \includegraphics[width=46mm,height=32mm]{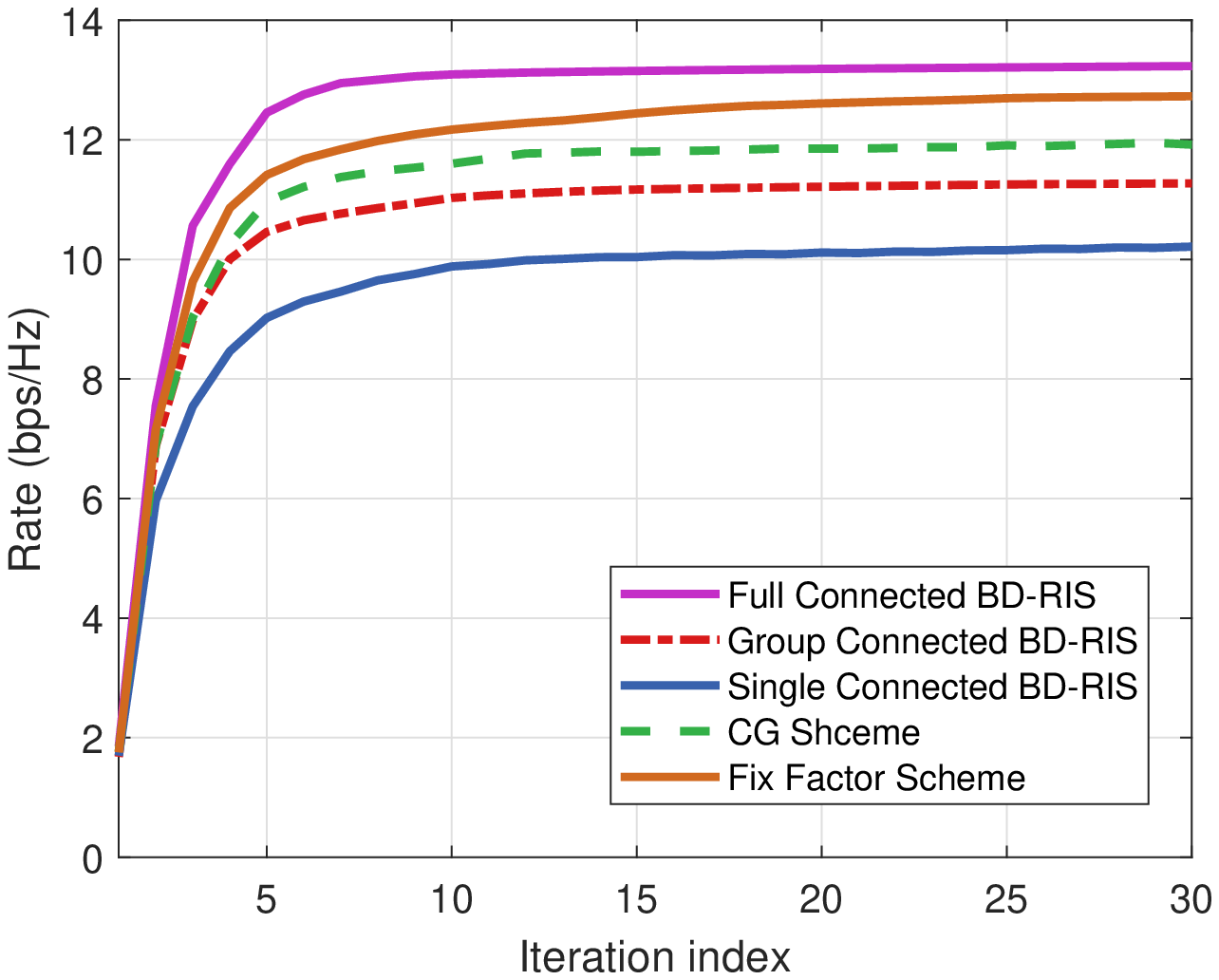}
                        \caption{Convergence of different schemes.}
                    \end{minipage}
                    \hfill
                    \begin{minipage}[b]{0.49\linewidth}
                        \centering
                        \includegraphics[width=46mm,height=32mm]{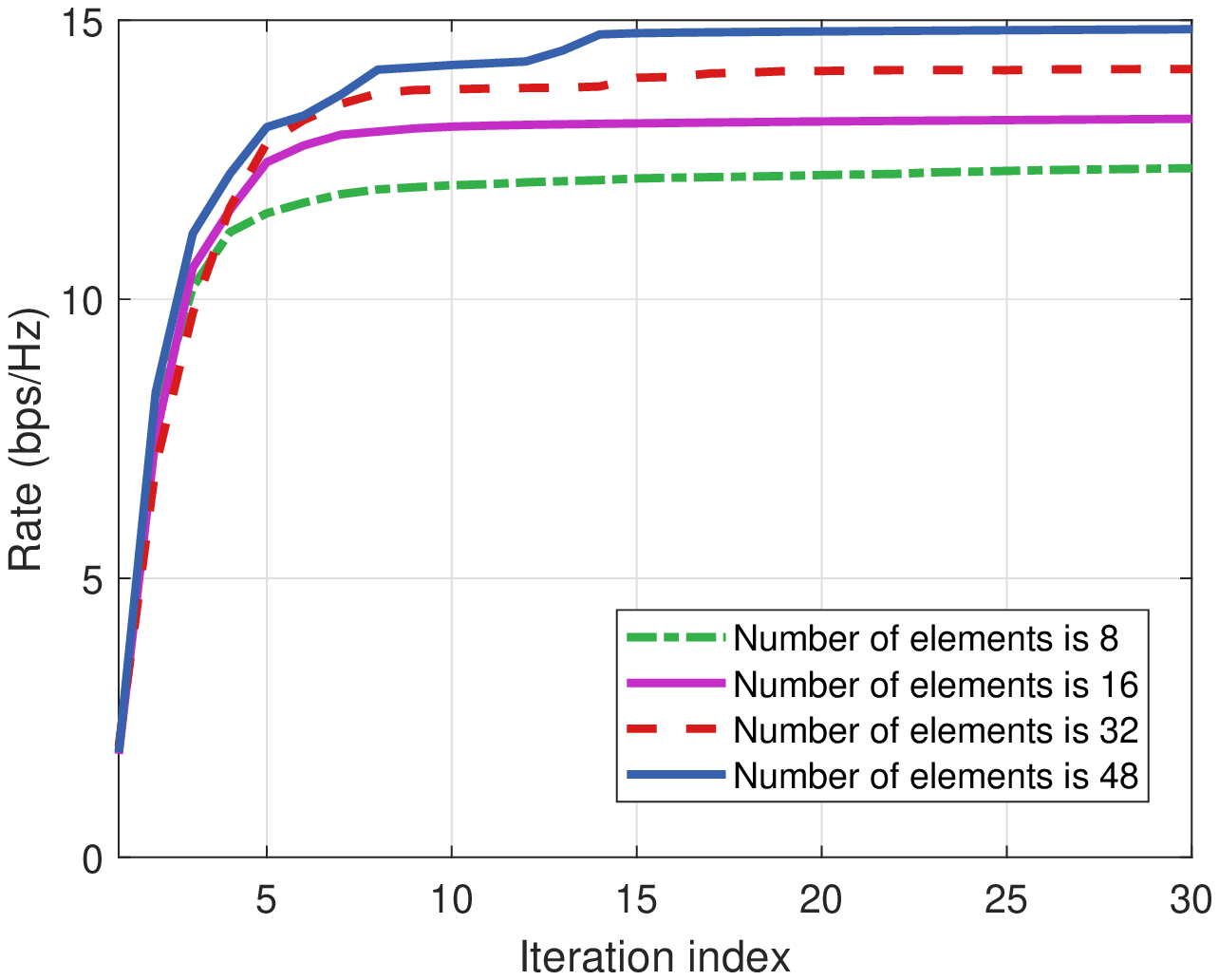}
                        \caption{Convergence of different number of BD-RIS elements.}
                    \end{minipage}
\end{figure}
This section provides numerical examples to verify the results. The sensing and communication noise powers are set to -80 dBm and -90 dBm, respectively. The BS has 8 antennas. The transmit power and the CRB threshold are 25 dBm and 0.001, respectively. The default number of elements in the BD-RIS, the number of groups, and the group size in the group connected mode are denoted by 16, 4, and 4, respectively. The distance of BS-RIS, RIS-UE, RIS-target are 40 m, 15 m, 18 m, respectively. The path-loss exponents for BS-RIS, RIS-UE, RIS-target, and BS-UE are 2.0, 2.0, 2.0, 3.0, respectively.

Fig. 2 shows the convergence of different schemes. The fully connected mode performs best, while the single connected is least efficient. Although the fully connected BD-RIS offers the best theoretical performance, its high complexity and cost make it less suitable when balancing performance, complexity, and cost.  Moreover, the proposed dynamic log-barrier method notably improves UEs' rate over the fix factor scheme and outperforms the conjugate gradient (CG) scheme \cite{BDRIS1}.

Fig. 3 and 4 illustrate the convergence of the proposed algorithm under varying numbers of BD-RIS elements and groups. In Fig. 3, the convergence becomes noticeably slower as the number of  elements increases. In Fig. 4, as the number of groups increases and each group size becomes smaller, the convergence slows down notably compared to the fully connected case, due to the group-wise optimization process.

\begin{figure}
	\centering
                \begin{minipage}[b]{0.49\linewidth}
                        \centering
                        \includegraphics[width=46mm,height=32mm]{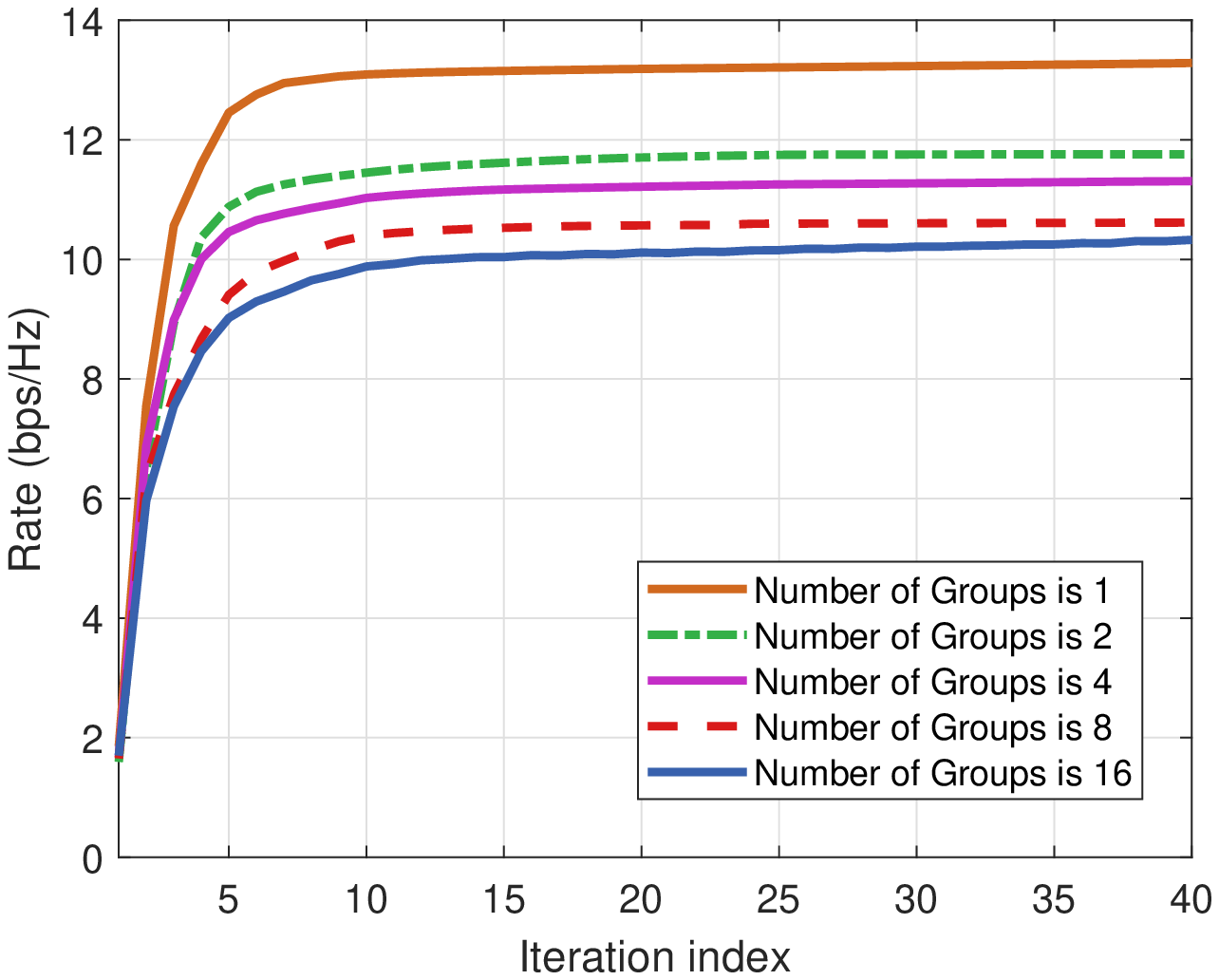}
                        \caption{Convergence of different number of BD-RIS gourps.}
                    \end{minipage}
                    \hfill
                    \begin{minipage}[b]{0.49\linewidth}
                        \centering
                        \includegraphics[width=46mm,height=32mm]{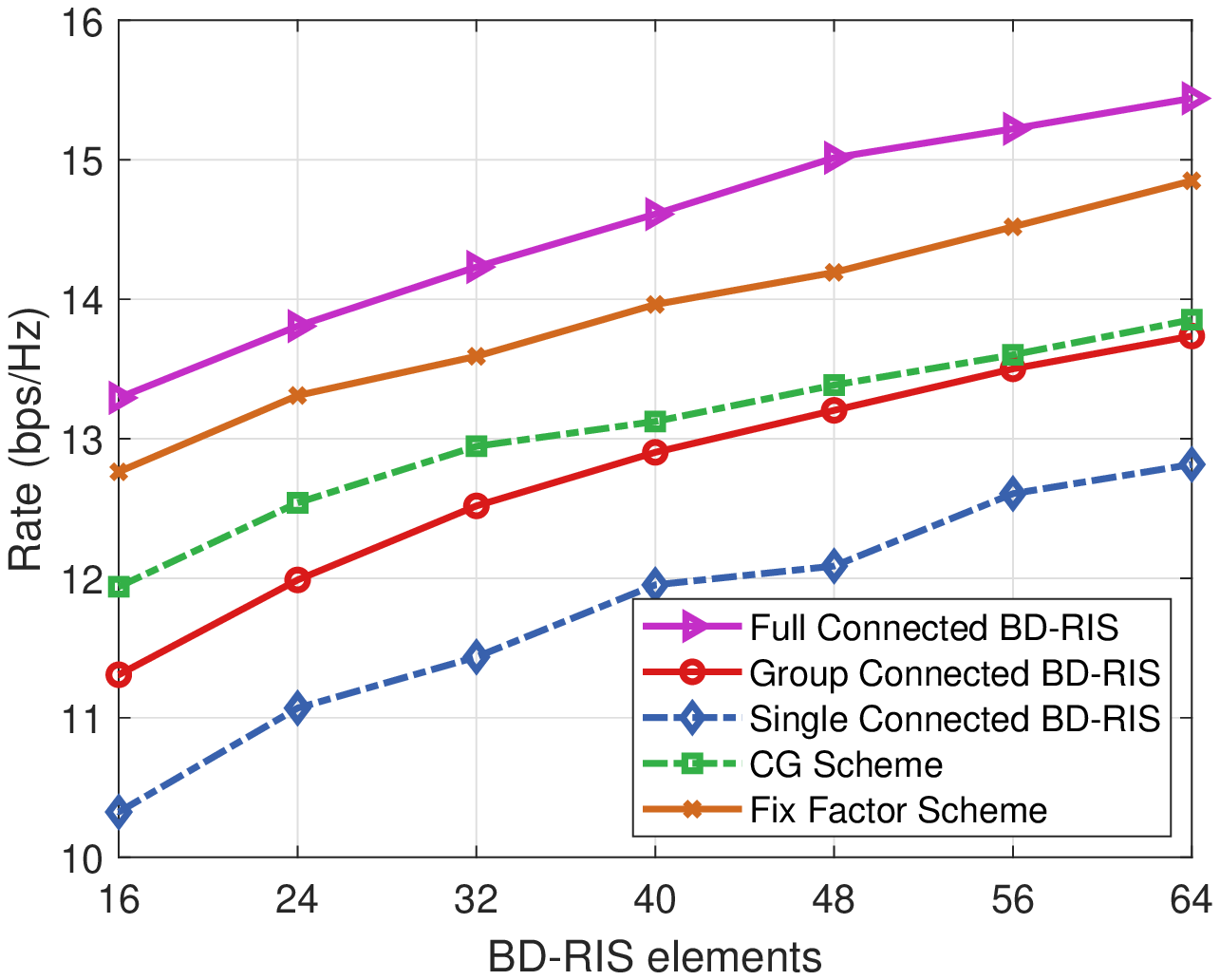}
                        \caption{Rate versus different number of BD-RIS elements.}
                    \end{minipage}
\end{figure}
\begin{figure}
	\centering
                \begin{minipage}[b]{0.49\linewidth}
                        \centering
                        \includegraphics[width=46mm,height=32mm]{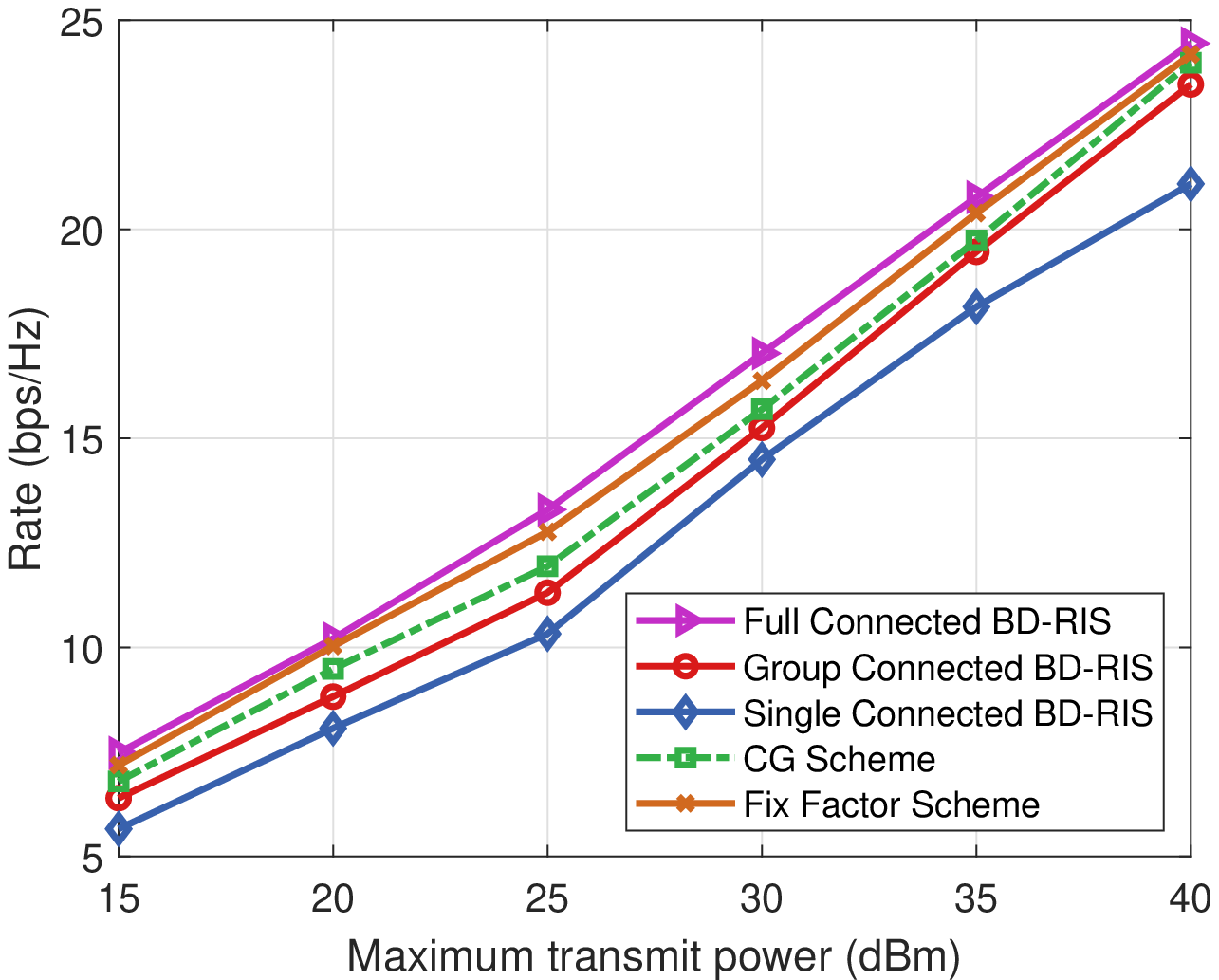}
                        \caption{Rate versus  transmit power.}
                    \end{minipage}
                    \hfill
                    \begin{minipage}[b]{0.49\linewidth}
                        \centering
                        \includegraphics[width=46mm,height=32mm]{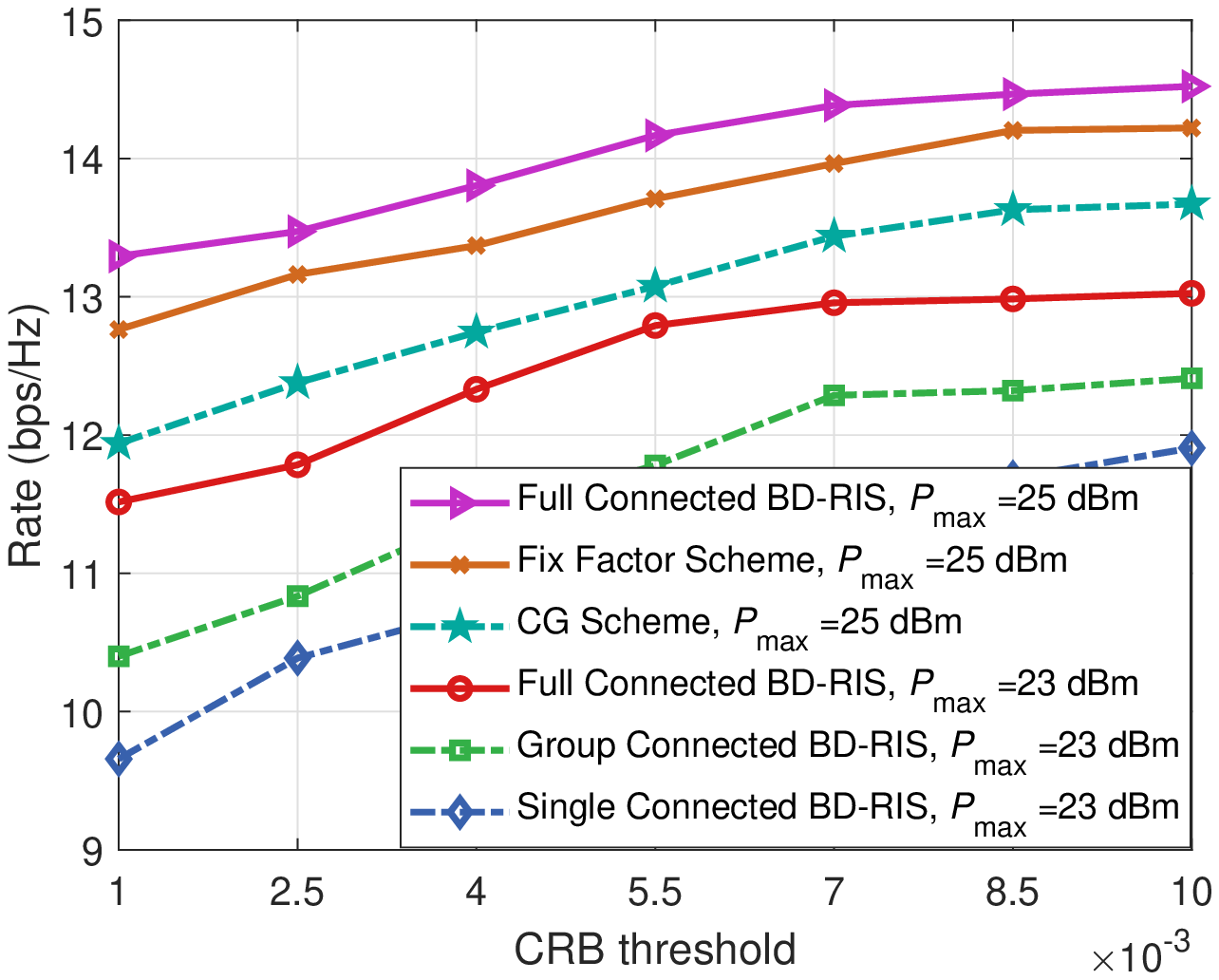}
                        \caption{Rate versus CRB threshold.}
                    \end{minipage}
\end{figure}
Fig. 5 presents that as the number of elements increases, the UE rate gradually improves across all three BD-RIS connection modes.  Furthermore, the proposed scheme consistently outperforms the CG scheme, with the effect of the dynamic log-barrier becoming more significant as element count grows.

Fig. 6 shows rate gains versus  transmit power.
The fully connected BD-RIS consistently outperforms conventional RIS as power increases, thanks to its internal structure. The proposed scheme also outperforms the CG scheme due to its flexible geodesic search and adaptive step size. Moreover, the dynamic log-barrier factor setting outperforms the fixed factor setting.

Fig. 7 exhibits the trade-off between the sensing and communication. A higher CRB threshold relaxes sensing demands, freeing resources for communication and boosting UEs' sum rate. Achieving the same rate with lower transmit power requires sacrificing some sensing performance. The proposed method efficiently allocates sensing and communication resources through joint optimization, meeting diverse sensing accuracy requirements while minimizing the impact on UEs' rate. As a result, UEs' rate remains stable across CRB thresholds, demonstrating an improved CRB–rate trade-off.

\section{Conclusion}
This letter presented an efficient joint design of the BS transmit beamforming matrix and the BD-RIS scattering matrix to balance the communication rate and the sensing accuracy in an ISAC system. Due to the CRB constraint and the unitary matrix constraint regarding to the BD-RIS scattering matrix, our considered problem belongs to constrained optimization on manifold.  We introduced a log-barrier-based adaptive Riemannian steepest ascent algorithm to solve this problem.  The proposed optimization framework can be applied to other BD-RIS aided systems where various types constraints about the scattering matrix exist.

\appendices
\section{Proof of Lemma 1} \label{Appendix_proof_I_xy_fq}

For convenience, let $R({{\bf{\Phi }}_\chi }) \!\!= \!\!\sum\nolimits_{k = 1}^K {R_k^1({{\bf{\Phi }}_\chi })\!\! -\!\! R_k^2({{\bf{\Phi }}_\chi })}  $, then
\begin{small} 
\begin{equation}
\begin{array}{l}
R_k^1({{\bf{\Phi }}_\chi }) = {\log _2}({\Lambda _{}}) = {\log _2}(\sum\nolimits_{i = 1}^K {|{\bf{h}}_{{\rm{BU}},k}^H({{\bf{\Phi }}_\chi }){{\bf{w}}_{{\rm{BS}},i}}{|^2}} \\ \qquad \qquad \quad{\rm{ + }}|{{\rm{d}}_{{\rm{TU}},k}} + {\bf{r}}_{{\rm{UE}},k}^H{{\bf{\Phi }}_\chi }{{\bf{r}}_{{\rm{Tar}}}}{|^2}{P_{{\rm{Tar}}}} + \sigma _{{\rm{UE}},k}^2),
\end{array}
\end{equation}
\end{small} 
\begin{small} 
\begin{equation}
\begin{array}{l}
R _k^2({{\bf{\Phi }}_\chi })  = {\log _2}({\Lambda _{\rm{}}}-{{{| {{\bf{h}}_{{\rm{BU}},k}^H( {{{\bf{\Phi }}_\chi }} ){{\bf{w}}_{{\rm{BS}},k}}} |}^2}}) .
\end{array}
\end{equation}
\end{small} 
The derivative of $R  _k^1({{\bf{\Phi }}_\chi })$ and $R  _k^2({{\bf{\Phi }}_\chi })$ are given by
\begin{equation}
\begin{array}{l}
\frac{{\partial R_k^1({{\bf{\Phi }}_\chi })}}{{\partial {\bf{\Phi }}_\chi ^{\rm{*}}}}{\rm{ \!=\! }}\frac{1}{{\ln 2}}\frac{{{{( {{{\bf{M}}_1} + {{\bf{M}}_5}} )}^T}}}{{\Lambda  }},

\frac{{\partial R_k^{\rm{2}}({{\bf{\Phi }}_\chi })}}{{\partial {\bf{\Phi }}_\chi ^{\rm{*}}}}{\rm{ = }}\frac{1}{{\ln 2}}\frac{{{{( {{{\bf{M}}_1} + {{\bf{M}}_5}} )}^T}}}{\Lambda - {{| {{\bf{h}}_{{\rm{BU}},k}^H( {{{\bf{\Phi }}_\chi }} ){{\bf{w}}_{{\rm{BS}},k}}} |}^2}}.
\end{array}
\end{equation}
Then the Euclidean gradient of $R ( {\bf{\Phi }_\chi} )$ is equal to (27).

For $\rho({\bf{\Phi }_\chi})$,  the differential is
\begin{equation}
{\rm{d}}\rho ({{\bf{\Phi }}_{\bf{\chi }}}) = {{\sigma _{{\rm{BS}}}^2{\rm{d}}\xi } \mathord{\left/
 {\vphantom {{\sigma _{{\rm{BS}}}^2{\rm{d}}\xi } {(2L{\xi ^2}\tau ({\rm{CR}}{{\rm{B}}_\theta }({{\bf{\Phi }}_{\bf{\chi }}}) - {\Delta _{\max }})}}} \right.
 \kern-\nulldelimiterspace} {(2L{\xi ^2}\tau ({\rm{CR}}{{\rm{B}}_\theta }({{\bf{\Phi }}_{\bf{\chi }}}) - {\Delta _{\max }})}}).
\end{equation}
where $\xi  = {\rm{tr}}({\bf{\dot h}}_{{\rm{TB}}}^{}({{\bf{\Phi }}_{\bf{\chi }}}){\bf{\dot h}}_{{\rm{TB}}}^H({{\bf{\Phi }}_{\bf{\chi }}})) - {\left| {{\rm{tr}}({\bf{U}})} \right|^{\rm{2}}}/{\rm{tr}}({\bf{V}})$. Let ${\xi _1}\!\! =\!\! {\rm{tr}}({\bf{\dot h}}_{{\rm{TB}}}^{}( {\bf{\Phi }_\chi}){\bf{\dot h}}_{{\rm{TB}}}^H( {\bf{\Phi }_\chi}))$, then the differential of ${\xi _1}$ is
\begin{equation}
{\mathop{\rm d}\nolimits} {\xi _1} =\! {\mathop{\rm tr}\nolimits} [ {{{{{{\bf{A}}_1}}  }}{\mathop{\rm d}\nolimits} ( {\bf{\Phi }_\chi} ) + {{\bf{A}}_2}{\mathop{\rm d}\nolimits} ( {{{\bf{\Phi }_\chi^*}}} )} ],
\end{equation}
where ${{{\bf{A}}_1}} = {\bf{\dot r}}_{{\mathop{\rm Tar}\nolimits} }^{}{\bf{\dot r}}_{{\mathop{\rm Tar}\nolimits} }^H{{{\bf{\Phi }}_\chi^H}}{\bf{G}}_{}^H{\bf{G}}$.
Let ${\xi _2} = {{{{| {{\mathop{\rm tr}\nolimits}( {\bf{\mathbf{U} }})} |}^2}} \mathord{/
 {\vphantom {{{{| {{\mathop{\rm tr}\nolimits} ( {\bf{\mathbf{U} }} )} |}^2}} {{\mathop{\rm tr}\nolimits} ( {\bf{\mathbf{V} }} )}}} 
 \kern-\nulldelimiterspace} {{\mathop{\rm tr}\nolimits} ( {\bf{\mathbf{V} }})}}$, then the differential of ${\xi _2} $ is 
\begin{equation}
        {\rm{d}}{\xi _2} = {{( {{\rm{d}}{\xi _3}{\mathop{\rm tr}\nolimits} ( {\bf{V}} ) - {\xi _3}{\rm{d}}[ {{\mathop{\rm tr}\nolimits} ( {\bf{V}} )} ]} )} \mathord{/
 {\vphantom {{( {{\rm{d}}{\xi _3}{\mathop{\rm tr}\nolimits} ( {\bf{V}} ) - {\xi _3}{\rm{d}}[ {{\mathop{\rm tr}\nolimits} ( {\bf{V}} )} ]} )} {{{[ {{\mathop{\rm tr}\nolimits} ( {\bf{V}} )} ]}^2}}}}
 \kern-\nulldelimiterspace} {{{[ {{\mathop{\rm tr}\nolimits} ( {\bf{V}} )} ]}^2}}},
\end{equation}
where ${\rm{d}}[ {{\rm{tr}}( {\bf{V}} )} ]
= {\rm{tr}}[ {{{\bf{D}}_1}{\rm{d}}( {\bf{\Phi }_\chi}) + {{\bf{D}}_2}{\rm{d}}( {{{\bf{\Phi }_\chi^*}}} )}]$ and ${{\bf{D}}_1} = {\bf{r}}_{{\mathop{\rm Tar}\nolimits} }^{}{\bf{r}}_{{\mathop{\rm Tar}\nolimits} }^H{{{\bf{\Phi }}_\chi^H}}{\bf{G}}_{}^H{\bf{G}}$.

Let ${\xi _3} = {\mathop{\rm tr}\nolimits} \left( {\bf{U}} \right){\left( {{\mathop{\rm tr}\nolimits} \left( {\bf{U}} \right)} \right)^*}$, then the differential of ${\xi _3} $ is 
\begin{equation}
        {\mathop{\rm d}\nolimits} {\xi _3}= {\mathop{\rm tr}\nolimits} [ {{{\bf{\Omega }}{{{\bf{\Phi }}_\chi^H}}{\bf{G}}_{{\mathop{\rm }\nolimits} }^H{\bf{G}}}{\rm{d}}( {{\bf{\Phi }}_\chi} ) + {{\bf{B}}_2}{\rm{d}}( {{{\bf{\Phi }_\chi^*}}} )} ].
\end{equation}

Substituting (43) into (42) yields
\begin{equation}
\begin{array}{l}
{\rm{d}}{\xi _2} = {\rm{tr}}\{ {( {\frac{{{{\bf{\Omega }}{{{\bf{\Phi }}_\chi^H}}{\bf{G}}_{{\mathop{\rm }\nolimits} }^H{\bf{G}}}}}{{{\rm{tr}}( {\bf{V}} )}} - \frac{{{\rm{tr}}( {\bf{U}} ){{( {{\rm{tr}}( {\bf{U}} )} )}^*}{{\bf{D}}_1}}}{{{{[ {{\rm{tr}}( {\bf{V}} )} ]}^2}}}} ){\rm{d}}( {{{\bf{\Phi }}_{\bf{\chi }}}} )} \\
\quad\qquad  { + ( {\frac{{{{\bf{B}}_2}}}{{{\rm{tr}}( {\bf{V}})}} - \frac{{{\rm{tr}}( {\bf{U}}){{( {{\rm{tr}}( {\bf{U}} )} )}^*}{{\bf{D}}_2}}}{{{{[ {{\rm{tr}}( {\bf{V}} )}]}^2}}}}){\rm{d}}( {{\bf{\Phi }}_{\bf{\chi }}^*} )} \}.
\end{array}
\end{equation}
Then ${\bf{\Sigma }}(\rho( {\bf{\Phi }_\chi}))$ can be represented as (28).

\end{document}